\documentclass[11pt,a4paper,twoside]{article}
\newcommand{\scr}{\scriptscriptstyle}

\newcommand{\muk}{\mu^{\scr +} K^0}
\newcommand{\epi}{e^{\scr +} \pi^0}
\newcommand{\et}{e^{\scr +} \eta}
\newcommand{\ek}{e^{\scr +} K^0}
\newcommand{\mpi}{\mu^{\scr +} \pi^0}
\newcommand{\mt}{\mu^{\scr +} \eta}
\newcommand{\ero}{e^{\scr +} \rho^0}
\newcommand{\eo}{e^{\scr +} \omega}

\newcommand{\mro}{\mu^{\scr +} \rho^0}
\newcommand{\mo}{\mu^{\scr +} \omega}
\newcommand{\nep}{\nu^{\scr C}_{\scr e} \pi^+}

\newcommand{\nmpi}{\nu^{\scr C}_{\scr \mu} \pi^+}
\newcommand{\nmk}{\nu^{\scr C}_{\scr \mu} K^+}
\newcommand{\nero}{\nu^{\scr C}_{\scr e} \rho^+}

\newcommand{\epin}{e^{\scr +} \pi^-}
\newcommand{\mpin}{\mu^{\scr +} \pi^-}
\newcommand{\eron}{e^{\scr +} \rho^-}
\newcommand{\mron}{\mu^{\scr +} \rho^-}
\newcommand{\nepn}{\nu^{\scr C}_{\scr e} \pi^0}
\newcommand{\nekn}{\nu^{\scr C}_{\scr e} K^0}
\newcommand{\nmpin}{\nu^{\scr C}_{\scr \mu} \pi^0}
\newcommand{\nmkn}{\nu^{\scr C}_{\scr \mu} K^0}
\newcommand{\neron}{\nu^{\scr C}_{\scr e} \rho^0}

\newcommand{\nmron}{\nu^{\scr C}_{\scr \mu} \rho^0}      

\newcommand{\neon}{\nu^{\scr C}_{\scr e} \omega}
\newcommand{\nmon}{\nu^{\scr C}_{\scr \mu} \omega}
\newcommand{\netn}{\nu^{\scr C}_{\scr e} \eta}
\newcommand{\nmtn}{\nu^{\scr C}_{\scr \mu} \eta}

\begin{document}
\def\NPB#1#2#3{Nucl. Phys. {\bf B} {\bf#1} (#2) #3}
\def\PLB#1#2#3{Phys. Lett. {\bf B} {\bf#1} (#2) #3}
\def\PRD#1#2#3{Phys. Rev. {\bf D} {\bf#1} (#2) #3}
\def\PRL#1#2#3{Phys. Rev. Lett. {\bf#1} (#2) #3}
\def\PRT#1#2#3{Phys. Rep. {\bf#1} C (#2) #3}
\def\ARAA#1#2#3{Ann. Rev. Astron. Astrophys. {\bf#1} (#2) #3}
\def\ARNP#1#2#3{Ann. Rev. Nucl. Part. Sci. {\bf#1} (#2) #3}
\def\MODA#1#2#3{Mod. Phys. Lett. {\bf A} {\bf#1} (#2) #3}
\def\NC#1#2#3{Nuovo Cim. {\bf#1} (#2) #3}
\def\ANPH#1#2#3{Ann. Phys. {\bf#1} (#2) #3}
\def\PTP#1#2#3{Prog. Th. Phys. {\bf#1} (#2) #3}

\begin{titlepage}
\thispagestyle{empty}
 
\begin{large}
\title{Large Quark Rotations Neutrino Oscillations and Proton Decay}
\end{large}


 \author{Yoav
  Achiman$^{a,b}$\footnote{e-mail:achiman@theorie.physik.uni-wuppertal.de}\quad
  \quad \\
  [1.5cm]
  {}$^a$School of Physics and Astronomy, Tel Aviv University, 69978 Tel Aviv,
  Israel\\
  {}$^b$Department of Physics,University of Wuppertal, D--42097 Wuppertal,
  Germany\\
  [1.5cm]} \date{ July 2001}

\maketitle

\begin{abstract} 
The large freedom in the SM fermionic mass matrices allows for large
RH and LH quark rotations. This is a natural possibility in view of the 
observed large leptonic mixing. Proton decay and especially its gauge 
mediated decay
is sensitive also to those mixing angles which are non-relevant in the SM.
A model with realistic mass matrices and large rotations is presented.
It is shown that the large leptonic mixing
leads to enhancement of the proton decay branching ratios involving muons.     
\end{abstract}
\end{titlepage}

\newpage
\noindent
Many {\em different} sets of fermionic mass matrices are consistent with the 
phenomenology of the SM.
This is due to the huge freedom in the fermionic rotations in the SM:\\
$V_{CKM}$---gives only the {\em difference} between the LH up and down quark 
mixing angles.\\
RH ROTATIONS---are unobservable in the SM (are equal to the LH rotations 
only if
the mass matrices are hermitian).\\
All quark mixing angles can be large, and in fact we know that large RH quark
rotations go naturally with the observed large LH leptonic 
mixing\cite{altarelli} in some GUT's.\\
In the SM the mass matrices are arbitrary hence the model must be extended 
to predict them (if not  for many other reasons). 
To find out what the  ``fundamental''
mass matrices in the extended theory are, one needs to know also the mixing
angles which are ``non relevant'' in the SM.\\
RH mixing can be observed via RH currents directly, or indirectly (if $W_R$
is very heavy) in: Baryon-asymmetry  induced via  Leptogenesis\cite{fy} 
(the relevant rotations are here RH !), Leptoquark interactions, Radiative 
corrections e.t.c.\\
In particular, the proton decay branching ratios are the best observables
to look for the ``non-relevant'' rotations. D=6 gauge mediated proton 
decays  involve all RH and LH mixing angles. They are also interesting
for the following reasons:\\ 
D=5 sparticle induced proton decays depend on yet unobserved sparticles and 
couplings and therefore involve many unknown parameters. This freedom 
however does not save SUSY SU(5) from being ruled out\cite{gn}. 
Also extensions of 
SU(5) and SO(10) must be carefully constructed and are yet on the verge 
of being excluded\cite{altarelli}~\cite{raby}. A natural alternative would 
be to suppress not only the 
D=4 but also the D=5 contributions (e.g. via a symmetry).\\
Gauge mediated decays involve only known coupling and masses and 
the predicted branching ratios are therefore much more reliable in this case.
They are the real test of GUTs, because D=5 decays are 
allowed also in non-GUTs as well. Also, there are recently quite a few 
models with $M_X$ lower than $10^{16}$ GeV~\cite{Mx} where D=6 proton decay 
can be observable in the near future.\\

To illustrate our point let me present a renormalizable SUSY SO(10) model 
with large mixing angles\cite{ar}. It is based on mass matrices with a 
non-hermitian Fritzsch texture induced via a $U(1)_F$ global family group.
The free parameters are chosen in such a way that all mixing angles will
be calculable in terms of the known fermionic masses and mixing.\\
The best fit for the mixing angles in the case of the favored LMA-MSW 
solution to the solar neutrino puzzle gives naturally large mixing angles.\\
\begin{enumerate}  
\item The Quark LH and RH mixing angles at the GUT scale:\\
\noindent
${\theta^u_L}_{12} = -0.077$,\quad ${\theta^u_L}_{23} = -1.48$,
\quad${\theta^u_L}_{13} = -4 \times 10^{-8}$.\\
${\theta^u_R}_{12} = -0.045$,\quad ${\theta^u_R}_{23} = -2.2\times
10^{-4}$,
\quad ${\theta^u_R}_{13} = -1.1 \times 10^{-3}$.\\
${\theta^d_L}_{12} = 0.15$,\quad ${\theta^d_L}_{23} = -1.44$,
\quad ${\theta^d_L}_{13} = 1 \times 10^{-5}$.\\
${\theta^d_R}_{12} = -0.33$,\quad ${\theta^d_R}_{23} = -3 \times 10^{-3}$,
\quad
${\theta^d_R}_{13} = 6 \times 10^{-2}$.\\

\item The mixing angles of the Charged Leptons:\\

\noindent
${\theta^\ell_L}_{12} = -1.17$,\quad ${\theta^\ell_L}_{23} = 1.44$,\quad
${\theta^\ell_L}_{13} = 0.0002 $.\\
${\theta^\ell_R}_{12} = 0.002$,\quad ${\theta^\ell_R}_{23} = -0.003$, \quad
${\theta^\ell_R}_{13} = 0.002 $.\\
\end{enumerate}                           
These were used in the calculation of the branching ratios for 
the proton and neutron decays. The following branching ratios result and 
are presented together with the case where the mixing is neglected in Table 1. 
\begin{table}\caption{Proton and neutron decay branching ratios}
\begin{center}
\centerline
\hfill
\begin{tabular}{|l|r|r||l|r|r|}
\hline
proton  & \% \ & \% & neutron & \% & \% \\
decay channel  & no mixing  & LA-MSW & decay channel & no mixing & LA-MSW \\
\hline
\hline
$p \; \rightarrow \; \epi$  
& 33.6 & 17.5 & 
$n \; \rightarrow \; \epin$
& 62.86 & 32.5 \\
$p \; \rightarrow \; \mpi$  
& -- & 16.1 & 
$n \; \rightarrow \; \mpin$
& -- & 30.0 \\
$p \; \rightarrow \; \ek$   
& -- & 4.6 & 
$n \; \rightarrow \; \eron$
& 9.7 & 5.0 \\
$p \; \rightarrow \; \muk$  
& 5.8 & 2.7 & 
$n \; \rightarrow \; \mron$
& -- & 4.6 \\
$p \; \rightarrow \; \et$   
& 1.2 & 0.6 & 
$n \; \rightarrow \; \nepn$
& 15.1 & 9.2 \\
$p \; \rightarrow \; \mt$   
& -- & 0.6 & 
$n \; \rightarrow \; \nekn$
& -- & 2.6 \\
$p \; \rightarrow \; \ero$ 
& 5.1 & 2.7 & 
$n \; \rightarrow \; \netn$
& 0.6 & 0.3 \\
$p \; \rightarrow \; \mro$  
& -- & 2.5 & 
$n \; \rightarrow \; \nmpin$
& -- & 5.1 \\
$p \; \rightarrow \; \eo$   
& 16.9 & 8.8 & 
$n \; \rightarrow \; \nmkn$
& 1.7 & 0.0 \\
$p \; \rightarrow \; \mo$   
& -- & 8.1 &  
$n \; \rightarrow \; \nmtn$
& -- & 0.2 \\
$p \; \rightarrow \; \nep$  
& 32.3 & 19.7 &  
$n \; \rightarrow \; \neron$
& 2.3 & 1.4 \\
$p \; \rightarrow \; \nmpi$ 
& -- & 10.9 & 
$n \; \rightarrow \; \neon$
& 7.7 & 4.7 \\
$p \; \rightarrow \; \nmk$  
& 0.1 & 0.2 &  
$n \; \rightarrow \; \nmron$
& -- & 0.8 \\
$p \; \rightarrow \; \nero$ 
& 4.9 & 3.0 & 
$n \; \rightarrow \; \nmon$
& -- & 2.6 \\
\hline
\end{tabular}
\hfill
\\
\end{center}
\end{table}
\vskip 0.1 cm
The branching ratios of the nucleon decay into muons are strongly enhanced
and are as large as the decay into $\epi$ .\\
The enhancement of the muon branching ratios is a unique feature of our
model because the decay mode $p \rightarrow e^+\pi^o$ is not negligible 
also in the $d=5$ induced decays\cite{bpw}.
In view of the fact that this enhancement is the effect of the large 
observed leptonic mixing on  the $d=6$ nucleon decay, we suggest 
that {\em the observation of a considerable rate for the decay $p \; 
\rightarrow \;\mu^+\pi^o$ will be a clear indication for a gauge 
mediated proton decay}.\\

One can say in general, that the branching ratios of the nucleon decay can
teach us about the ``fundamental'' mass matrices as they depend on all mixing
angles. The present huge freedom in the mass matrices would then be strongly 
restricted and one could better understand the origin of the fermionic masses.

\end{document}